%% file: main.tex
\PassOptionsToPackage{table}{xcolor}
\documentclass[color=00935A]{seedstyle}

\input{preamble}

\title{\raisebox{-0.1cm}{\includegraphics[height=0.6cm]{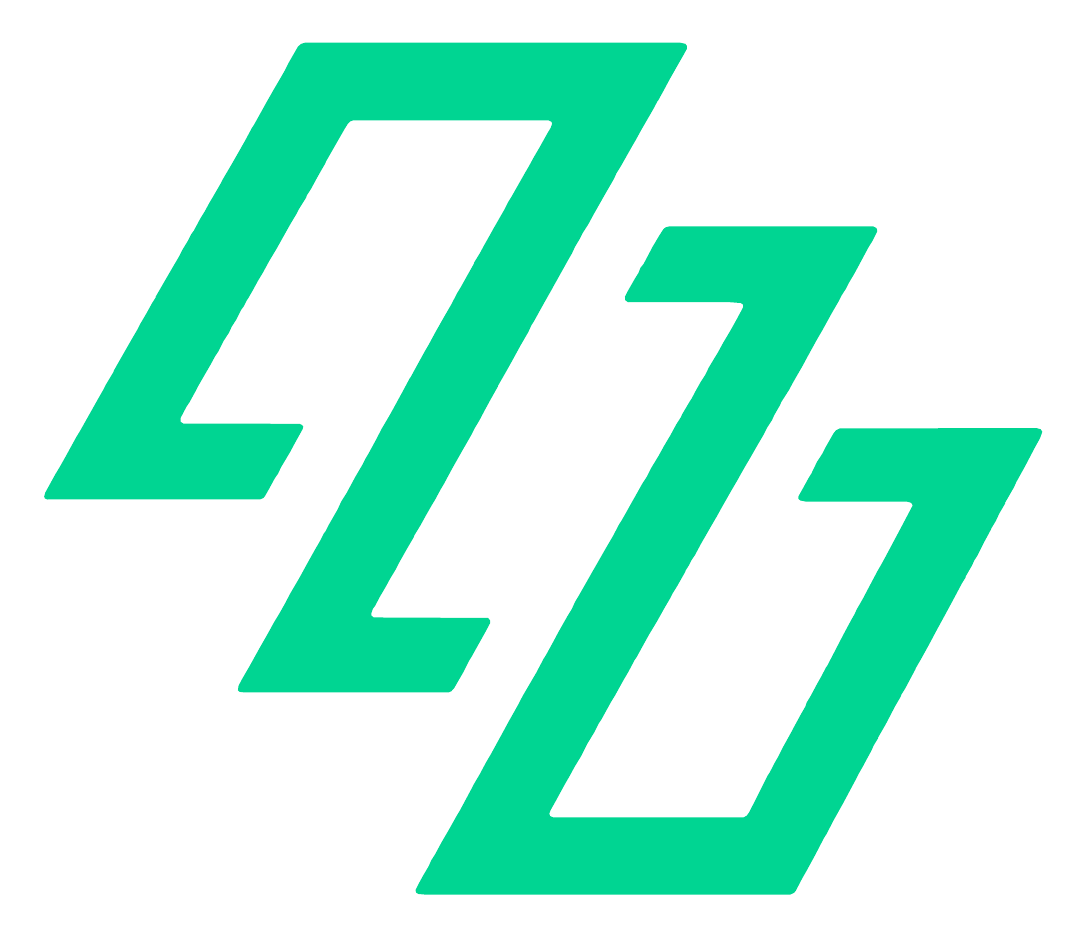}} Micius-Flash:\\ Lightweight Agentic Intelligence for Engineering}

\vspace{-1.5cm}
\author{Micius Team}
\vspace{-3cm}

\abstract{\input{sections/00_abstract}}

\date{\today}

\vspace{-0.4cm}

\begin{document}
\vspace*{-0.8cm}
\maketitle

\begin{center}
    \includegraphics[
        width=\textwidth,
        height=0.32\textheight,
    ]{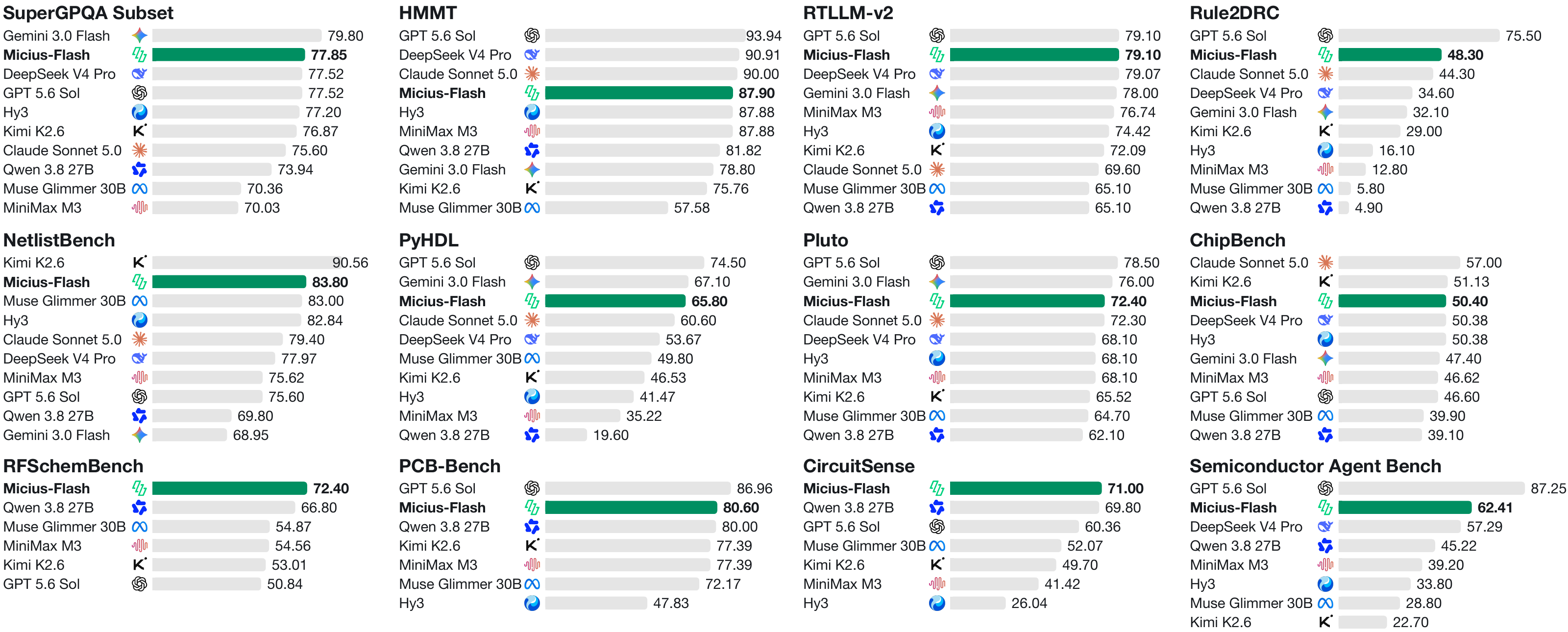}
    \captionof{figure}{
        Micius-Flash performance across benchmarks
        (see Section~\ref{sec:Evaluation} for details).
    }
    \label{fig:result}
\end{center}

\input{sections/01_introduction}

\input{sections/02_data_curation}

\input{sections/03_Post_training}

\input{sections/04_Evaluation}

\input{sections/05_Micius_harness}

\input{sections/06_Case_study}

\input{sections/07_conclusion}
\input{sections/08_Contributions}

\bibliographystyle{plainnat}
\bibliography{references}



\end{document}

%% file: preamble.tex
\usepackage{wrapfig}
\usepackage{dashrule}
\usepackage{mathtools}
\usepackage{pifont}
\usepackage{tabularx}\usepackage{makecell}
\usepackage{float}
\usepackage{flafter}
\usepackage{placeins}
\usepackage{booktabs}
\usepackage{tabularx}
\usepackage{array}
\usepackage{graphicx}
\usepackage{caption}
\usepackage{float}
\usepackage[table]{xcolor}
\graphicspath{{images/}}
\definecolor{taskblue}{HTML}{EAF1FA}
\definecolor{taskteal}{HTML}{E8F3EF}
\definecolor{taskpurple}{HTML}{F0ECF7}
\definecolor{taskgold}{HTML}{FAF1DF}
\definecolor{taskrose}{HTML}{F8EBEE}

\usepackage{needspace} 

\usepackage{pdflscape}
\usepackage{longtable}

%% file: sections/00_abstract.tex
We introduce Micius-Flash, a 27-billion-parameter model for engineering with native multimodal input, deep reasoning and a 256K-token context window. We design the Micius data pipeline as a knowledge-tree-driven data flywheel comprising three stages: knowledge tree construction and expansion, data synthesis, and quality validation with hard-example augmentation. From this data, we train expert models for semiconductor knowledge and reasoning, hardware coding, and semiconductor multimodal understanding, along with agentic experts. Subsequently, we employ Multi-Teacher On-Policy Distillation (MOPD) to supervise training on sequences generated by Micius itself, thereby integrating the capabilities of these expert models into a unified model. On more than 20 datasets with about 15K records, Micius-Flash averages 64.2\% on hardware code benchmarks, 15.0 points above a same-size baseline. It scores 62.4\% on a semiconductor-specific agent benchmark, exceeding the same-size baseline by 17.2 points. We further develop Micius Harness for industrial applications, connecting the model to specialized tools, including EDA tools and KiCad, as well as execution environments. Case studies in PCB design, chip design workflow automation, and software development demonstrate the model's ability to use tool execution feedback for multi-step planning, code debugging, and iterative verification.
We release Micius as an open-source model to advance the development of agents for science and engineering.

%% file: sections/01_introduction.tex
\section{Introduction}
\label{sec:introduction} 

Recent advances in large language models (LLMs) have expanded their capabilities in reasoning, code generation, multimodal understanding, and interaction with external tools \cite{anthropic2026opus5,google2026gemini37flash,openai2026gpt6astra,moonshot2026kimik3,qwen2026qwen38max}. Research on software agents further demonstrates the importance of interfaces and execution environments for complex, multi-step tasks \cite{yang2024swe,gao2026engiworld,wang2025openhands}. These developments create new opportunities for scientific research and engineering, where models must work with specialized knowledge, heterogeneous information, and professional software beyond conventional text-based interactions.

Semiconductor research and engineering exemplify these demands, spanning device physics, manufacturing processes, circuit design, and verification. Tasks may involve interpreting schematics, generating hardware code or EDA scripts, and refining designs using simulation feedback. Domain-adapted models show the value of specialized training \cite{liu2023chipnemo,liu2024rtlcoder,wu2024chateda}, while benchmarks expose limitations in hardware generation, circuit understanding, and realistic design tasks \cite{verilogeval,rtllm,circuitsense,chipbench}. Three challenges remain. First, engineering knowledge is fragmented and insufficiently covered by general-purpose training data, motivating systematic domain data collection, synthesis, and quality control \cite{gururangan2020don,wang2023self}. Second, engineering tasks require complementary capabilities that are difficult to consolidate within a single model without interference or degradation of general-purpose performance \cite{ma2026mopd}. Third, applying model capabilities to real engineering workflows requires dedicated execution infrastructure. Specialized tools have distinct interfaces and runtime dependencies, while multi-stage tasks require coordinated tool access, environment configuration, and executable workflow. 

In this report, we introduce Micius-Flash, a 27B-parameter multimodal language model for scientific and engineering tasks, with semiconductor engineering as its primary domain of specialization and evaluation. Micius supports native multimodal understanding, extended reasoning, and a context window of up to 256K tokens. We develop a knowledge-tree-driven data flywheel that organizes semiconductor knowledge, guides targeted data collection and synthesis, and uses quality validation and evaluation feedback to address coverage gaps. We independently fine-tune multiple specialized teacher models for complementary domain-specific and agentic capabilities, and initialize the student through supervised warmup. We then consolidate these capabilities through multi-teacher on-policy distillation, with token-level supervision on trajectories generated by the evolving student policy \cite{agarwal2024onpolicydistillationlanguagemodels}. In addition, we incorporate engineering examples and tool-use trajectories verified through compilation, simulation, and sandbox execution into specialist training, drawing on execution outcomes rather than text-level supervision alone.

We evaluate Micius-Flash on more than 20 benchmarks comprising approximately 15,000 instances across semiconductor knowledge, hardware coding, multimodal understanding, scientific reasoning, and agentic tasks. On 12 hardware coding benchmarks, Micius averages 64.2\%, outperforming the same-size general-purpose baseline by 15.0 points; on a semiconductor-specific agent benchmark, it scores 62.4\%, a gain of 17.2 points. The model also remains competitive on general-purpose knowledge and reasoning evaluations, including SuperGPQA and MMLU-Pro \cite{supergpqa,mmlupro}. In addition, we develop Micius Harness, an engineering workbench connecting the model to professional tools, including EDA software and KiCad, and to configured execution environments. It supports scoped tool access, multi-stage task execution, and engineering artifact inspection. Case studies in chip design automation, PCB engineering, industrial software development, and scientific experimentation illustrate how Micius uses tool feedback to refine intermediate solutions and produce verifiable engineering artifacts.

\begin{itemize}
    \item We develop a knowledge-tree-driven data flywheel for systematic semiconductor knowledge organization, targeted data synthesis, and iterative refinement guided by evaluation feedback.

    \item We consolidate complementary capabilities from multiple specialized teachers into a unified 27B model through multi-teacher on-policy distillation, integrating domain expertise, multimodal understanding, coding, and tool use.

    \item We develop Micius Harness to bridge language models and real engineering workflows through professional tool integration and executable environments, supported by execution-verified training and representative engineering case studies.
\end{itemize}

%% file: sections/02_data_curation.tex
\section{Data Curation}

\label{sec:Data_curation}

We design the Micius data pipeline as a knowledge-tree-driven data flywheel, organizing data production into three stages: knowledge tree construction and expansion, data synthesis, and quality validation with hard-example feedback. The knowledge trees capture the knowledge hierarchy, conceptual relationships, and dependencies within the semiconductor domain, linking source data, synthetic samples, and model diagnostics to a unified knowledge structure. Within this framework, each data production cycle begins with explicit knowledge coverage requirements and identified capability gaps. Targeted data collection and synthesis augment the training data, while quality validation and model evaluation results guide updates to the knowledge structure and sampling strategies, forming a continuously iterating, closed-loop data production process.

\begin{figure}[htbp]
    \centering
    \includegraphics[width=\linewidth]{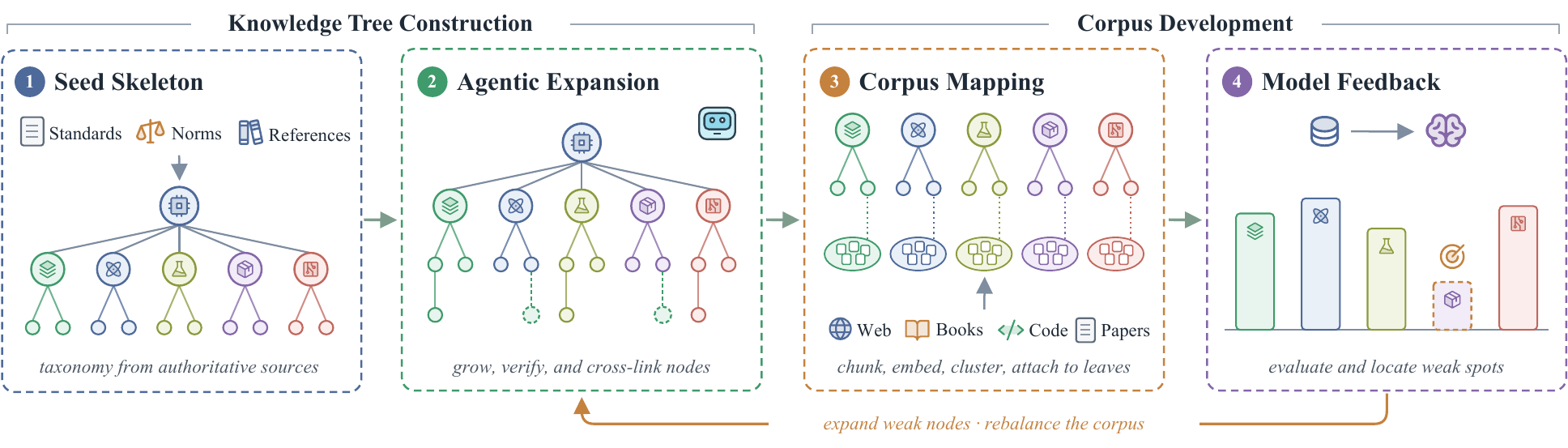}
    \caption{Knowledge-tree-driven data flywheel for semiconductor training data. We build and extend a domain knowledge tree, then map source corpora onto its nodes. Evaluation results point to the nodes where the model is weak, and these guide the next round of data synthesis and refinement.}
    \label{fig:knowledge_graph}
\end{figure}

\subsection{Knowledge Tree Construction}
We first construct knowledge trees based on authoritative standards, normative documents, and core reference materials in the semiconductor domain. We progressively refine these trees from domains and subdomains to concepts and methods, establishing a clearly defined knowledge hierarchy. This structure provides a unified organizational basis for data collection and supports subsequent coverage analysis of synthetic data and localization of model capability gaps.

Building on the initial knowledge structure, agents perform document retrieval, concept extraction, relationship identification, and recursive expansion around existing nodes. Before incorporation, candidate nodes are aligned with existing concepts and deduplicated, with checks on conceptual granularity, parent--child relationships, and supporting sources. Further subdivision stops when a node represents a well-defined, self-contained unit of knowledge. Beyond the tree hierarchy, we add dependencies and cross-branch relationships between concepts, forming a knowledge graph that supports knowledge retrieval, compositional sampling, and the mapping of diagnostic feedback to relevant knowledge nodes.

\subsection{Data Synthesis}
Synthesis is indexed by the knowledge trees. Each sample is generated from a knowledge node or a local subgraph together with its source evidence, and keeps a mapping back to those nodes for coverage analysis and hard-example feedback.

\noindent\textbf{Reasoning and Knowledge.}
We sample single nodes, parent--child pairs, and cross-branch combinations, and control difficulty by the number of nodes and reasoning hops. Several strong models answer each question independently, and we keep only samples whose answers agree with the reference answer and the source evidence.

\noindent\textbf{Coding.}
We use knowledge nodes, GitHub repositories, and programming manuals as seeds to construct tasks such as specification-to-RTL generation, code completion, bug fixing, and testbench generation. Candidates are verified by compilation and simulation, and only passing samples are kept.

\noindent\textbf{Visual.}
Images are rendered from schematic codes, waveform codes, layout codes and RTL simulation codes, and are linked to knowledge nodes through their captions and surrounding text. We generate questions on component identification, circuit structure, waveform and chart reading, and reasoning that combines the image with domain knowledge, and discard samples that a text-only model can answer without the image.

\subsection{Data Quality Validation}
We evaluate synthetic data along three dimensions: sample diversity, difficulty distribution, and question--answer quality. Sample diversity concerns coverage across domains, knowledge nodes, combinations of knowledge, and task formats. Difficulty distribution concerns the balance between basic and integrative tasks, as well as the alignment between task difficulty and the model’s current capabilities. Question--answer quality concerns the completeness of task conditions, task solvability, answer correctness, and consistency with supporting source evidence. These checks remove duplicate, ambiguous, and insufficiently supported samples, yielding a dataset suitable for training.

Beyond sample-level quality checks, we assess the training utility of synthetic data through small-scale supervised fine-tuning (SFT) and independent evaluation. Under consistent training configurations and evaluation protocols, we analyze model performance across domains, knowledge nodes, and difficulty levels to identify gaps in data coverage, imbalances in difficulty, and persistent capability limitations. Sample validation and training-based evaluation provide feedback on data reliability and training effectiveness, respectively, jointly guiding the next data production cycle.

For domains with weak evaluation performance, we map failed examples back to the corresponding knowledge nodes, relationships, or local subgraphs and analyze their primary causes. If failures stem from insufficient knowledge coverage or overly coarse node granularity, we further expand the relevant leaf nodes and retrieve targeted resources. If they arise from inadequate coverage of knowledge combinations or task difficulty levels, we adjust subgraph sampling and synthesis strategies. If they result from missing evidence, ambiguous questions, or incorrect answers, we prioritize data correction. Reviewed hard examples and their derived samples are incorporated into subsequent iterations. Through this process, model feedback continuously drives knowledge graph expansion and updates to the data distribution, ensuring that dataset growth is accompanied by broader knowledge coverage, greater task diversity, and expanded coverage of informative hard examples.

\subsection{Agentic Environments and Task Synthesis}

\begin{figure}[ht]
    \centering
    \includegraphics[width=\linewidth]{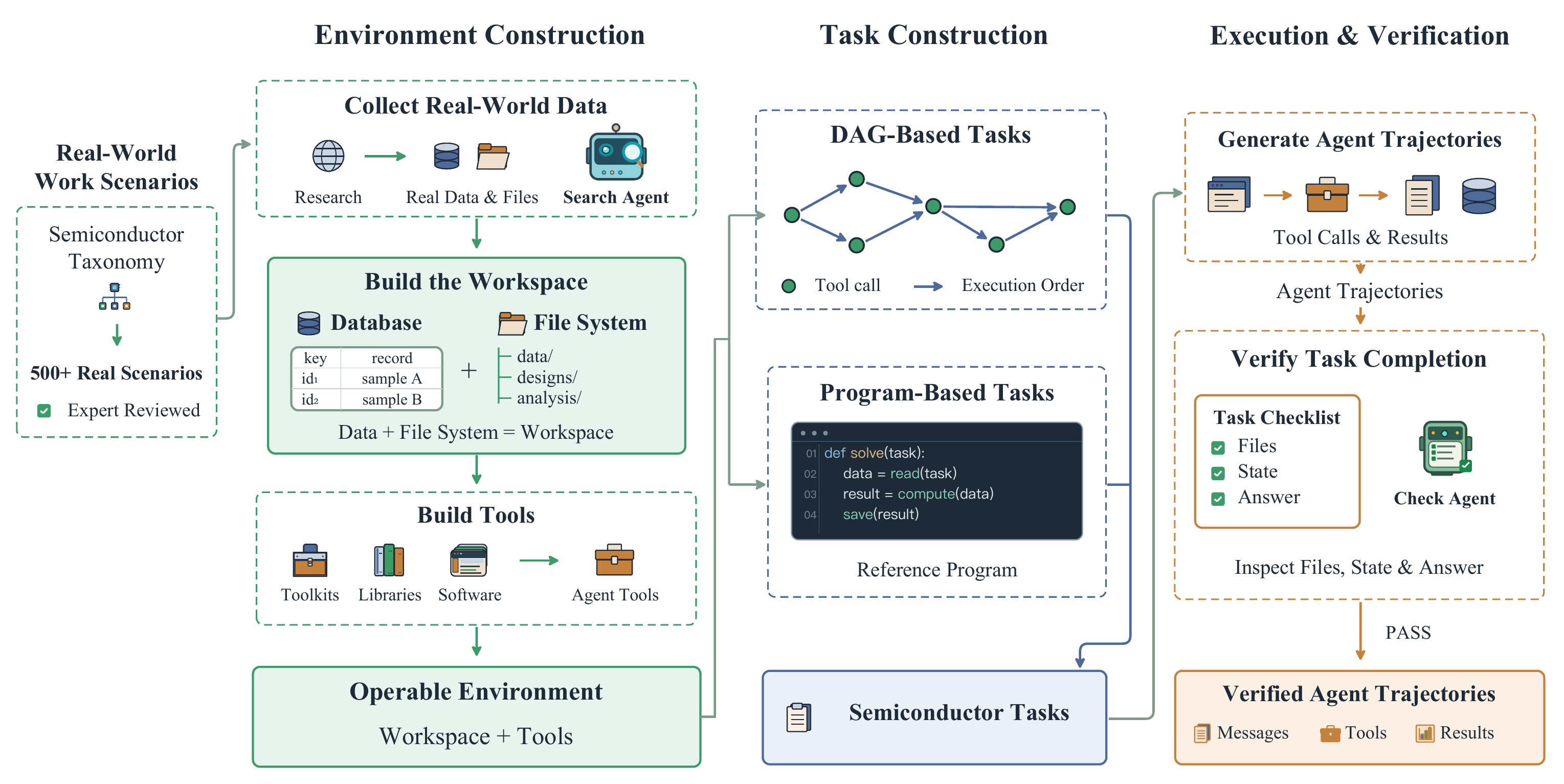}
    \caption{Overview of the agentic environment and task synthesis pipeline, from real-world scenarios to environment construction, task generation, and trajectory verification.}
    \label{fig:agent-task-pipeline}
\end{figure}

Agentic training data for scientific and industrial workflows remains scarce. Synthesizing such data requires both production at scale and fidelity to practical requirements. To address this challenge, we design a data synthesis pipeline for the semiconductor domain that converts authentic work scenarios reviewed by experts into interactive environments and uses them to generate tasks involving multiple steps. By retrieving data and files from authentic sources, the pipeline grounds the environment in practical workflows. It also supports the expansion of operational capabilities across scenarios through the reuse and composition of modular tool components. The resulting tasks cover data management as well as analysis and design work in the semiconductor domain. Figure~\ref{fig:agent-task-pipeline} provides an overview of the complete pipeline, including environment construction, task construction, and execution and verification.

We first refer to authoritative taxonomies and definitions in the semiconductor domain to develop a detailed scenario classification system covering materials and device research and development, device and physics simulation, and chip design and tapeout. Based on this classification, we collect more than 500 authentic work scenarios in the semiconductor domain and have domain experts assess their authenticity and usability. These scenarios form a seed set for subsequent environment construction and task synthesis.

For each seed scenario, the pipeline conducts further research to supplement its context and the business information needed to guide data retrieval and environment construction. A Search Agent then retrieves database resources and files relevant to the scenario. The pipeline imports data suited to database storage into a database and organizes files available for direct operation in a file system, forming a workspace that supports both data queries and file operations.

Using this workspace, the pipeline selects and combines Python packages, tool libraries, and interfaces to specialized software according to the operations required by each scenario. It then wraps these components as tools that the agent can call. By reusing existing components and combining operational interfaces as needed, the pipeline supports the tool requirements of different scenarios and builds an interactive agent environment containing both data resources and operational tools.

Within these environments, we construct tasks programmatically using directed acyclic graphs (DAGs) to represent operations involving multiple steps and their dependencies. The task construction process uses the environment information available to it to design tasks, while the executing agent receives only the task objective and some relevant conditions. The agent must use tools to retrieve and analyze information in the environment and complete each task through successive steps. This information asymmetry allows the tasks to evaluate proactive information gathering, multistep reasoning, and tool coordination. It also supports the construction of tasks with varied objectives and operational dependencies from the same environment.

To evaluate the generalizability and practical usability of our data synthesis pipeline beyond the semiconductor domain, we selected game development, a popular domain for agentic applications, and generated agentic data for this domain. We further evaluated the pipeline's outputs on additional game development tasks. The results demonstrate the effectiveness of the generated data and provide empirical evidence that the pipeline can be applied to agentic applications beyond the semiconductor domain.

%% file: sections/03_Post_training.tex
\section{Post-training}
\label{sec:Post_training}

\begin{figure*}[t]
    \centering
    \includegraphics[
        width=\textwidth,
        trim={0bp 90bp 0bp 110bp},
        clip]{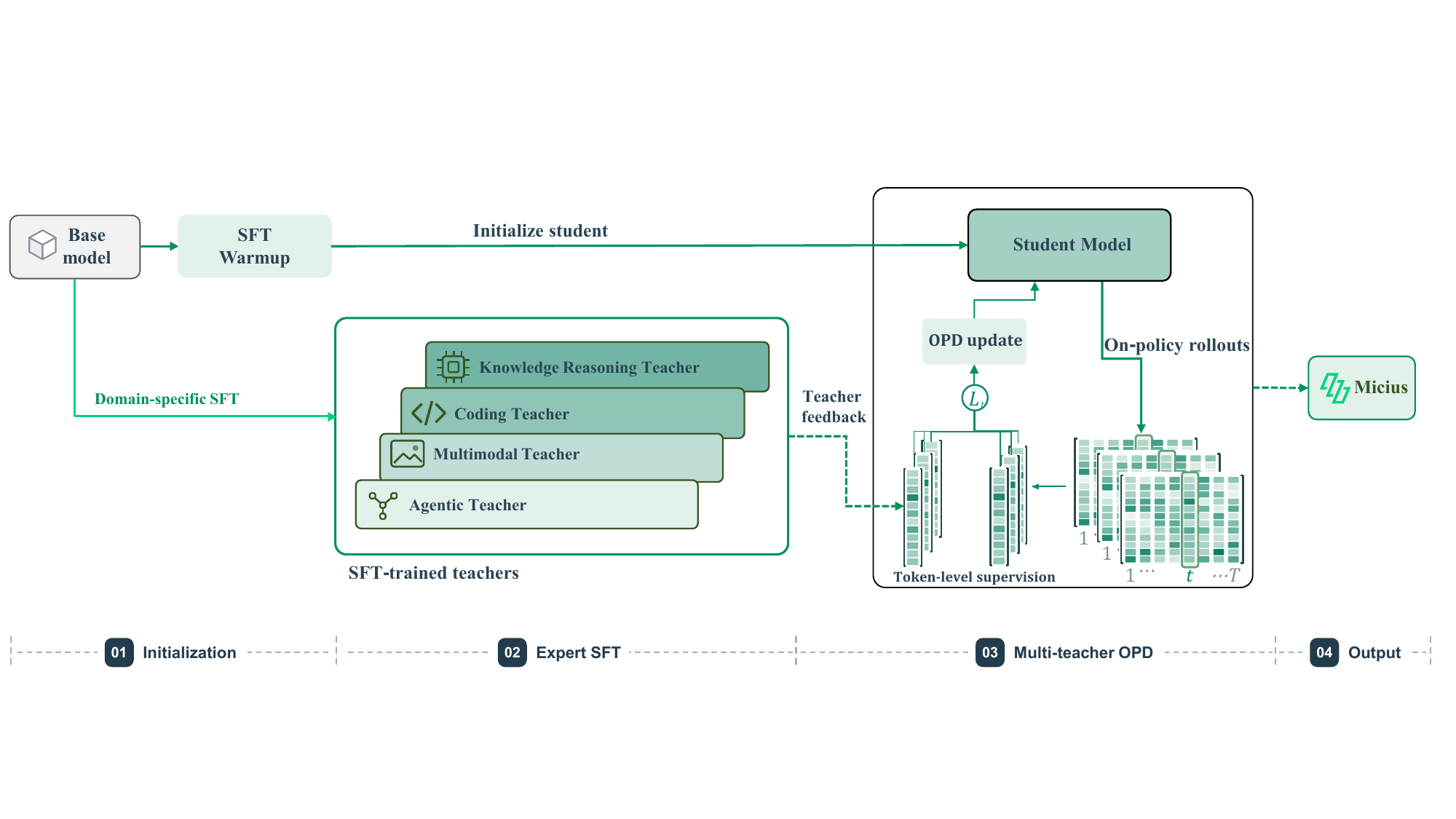}
    \caption{Overview of the Micius-Flash post-training pipeline. Starting from the base model, SFT produces four specialized teachers for semiconductor knowledge question answering and reasoning, coding, multimodal question answering, and agentic tasks, while initializing the student through supervised warmup. During MOPD, the student generates on-policy trajectories and receives token-level supervision from the teacher assigned to each task category, iteratively consolidating their capabilities into a unified model.}
    \label{fig:training_pipeline}
\end{figure*}

As illustrated in Figure~\ref{fig:training_pipeline}, the post-training pipeline of Micius-Flash consists of two stages: Supervised Fine-Tuning (SFT) and Multi-Teacher On-Policy Distillation (MOPD)~\cite{ma2026mopd}. SFT produces four teachers specialized in different task categories within the semiconductor domain and initializes the student policy through supervised warmup. MOPD then consolidates the teachers' complementary capabilities into a unified student through token-level distributional supervision on trajectories generated by the current student policy, yielding the final Micius model.

\subsection{Supervised Fine-Tuning}
\label{sec:sft}

The SFT stage comprises specialized teacher training and student policy initialization. Starting from the base model, we independently fine-tune four teachers on data for their respective task categories within the semiconductor domain. The knowledge and reasoning teacher focuses on domain-specific knowledge question answering and reasoning. The coding teacher specializes in semiconductor-related coding tasks. The multimodal teacher handles semiconductor question answering involving visual and textual information. The agentic teacher focuses on semiconductor agentic tasks, including tool use. Each teacher is optimized for its assigned tasks to strengthen specialized knowledge and task-solving capabilities, providing complementary sources of supervision for subsequent distillation. For the student, we perform supervised warmup on the base model using the SFT data and use the resulting checkpoint to initialize MOPD. Building on the base model's existing capabilities, this stage further strengthens semiconductor knowledge and task-solving abilities, providing a domain-adapted initial policy for multi-teacher distillation.

\subsection{Multi-Teacher On-Policy Distillation}
\label{sec:mopd}

Following SFT, we apply MOPD to consolidate the capabilities of the four specialized teachers into a single student model~\cite{mimo2026v2flash}. During training, the current student policy generates responses or interaction trajectories, and each trajectory is routed to the teacher associated with its task category for token-level distributional supervision~\cite{agarwal2024onpolicydistillationlanguagemodels}. We use a forward KL objective that retains the teacher's top-32 token probabilities and aggregates the remaining probability mass into a single tail category. Let $p_t(v)=p_{T_c}(v\mid h_t)$ and $q_t(v)=\pi_\theta(v\mid h_t)$ denote the probabilities assigned to token $v$ by the teacher $T_c$ for task category $c$ and the student, respectively, given the current context $h_t$. With $\mathcal{S}_t$ denoting the teacher's top-32 token set, the per-token distillation loss is
\begin{equation}
\label{eq:mopd_token_loss}
\ell_t =
\sum_{v\in\mathcal{S}_t}
p_t(v)\log\frac{p_t(v)}{q_t(v)}
+
(1-m_{T,t})\log\frac{1-m_{T,t}}{1-m_{S,t}},
\end{equation}
where $m_{T,t}=\sum_{v\in\mathcal{S}_t}p_t(v)$ and $m_{S,t}=\sum_{v\in\mathcal{S}_t}q_t(v)$. All probabilities are normalized over the full vocabulary. The loss is the forward KL divergence between the coarsened distributions obtained by retaining the top-32 tokens and aggregating all remaining tokens into a single tail category. Each trajectory receives supervision only from its assigned teacher. For batches containing multiple task categories, we average the loss over valid token positions within each category and combine the category losses using their respective weights. As the student policy evolves, trajectories are regenerated online, enabling the teachers' specialized knowledge and task-solving behaviors to be transferred in contexts visited by the student.

%% file: sections/04_Evaluation.tex
\section{Evaluation}
\label{sec:Evaluation}

\subsection{Benchmarks}

To supplement existing benchmarks' coverage of semiconductor
knowledge and end-to-end task execution, we construct
SemiBench, Semiconductor Agent, and PCB Agent.
These benchmarks assess complementary capabilities.
SemiBench evaluates semiconductor-domain knowledge and reasoning.
Semiconductor Agent and PCB Agent evaluate agentic task solving
in semiconductor and PCB engineering, respectively.
Our evaluation combines these with existing benchmarks across
four capability categories: reasoning and knowledge, coding,
vision, and agentic capabilities.

\noindent\textbf{Reasoning and Knowledge.}
We evaluate semiconductor knowledge and reasoning using
SemiBench and four selected subsets of SuperGPQA~\cite{supergpqa}:
circuits, electromagnetics,
microelectronics, and physics.
We evaluate general mathematical reasoning using
HMMT~\cite{hmmt}
and broader knowledge using MMLU-Pro~\cite{mmlupro}.

\noindent\textbf{Coding.}
We evaluate hardware-oriented code generation and engineering
programming using CVDP-static~\cite{cvdp},
VerilogEval-v2~\cite{verilogeval},
RTLLM-v2~\cite{rtllm},
ArchXBench~\cite{archxbench},
NetlistBench~\cite{netlistbench},
VHDLBench~\cite{vhdlbench},
Pluto~\cite{pluto},
NotSoTiny~\cite{notsotiny},
Rule2DRC~\cite{rule2drc},
PyHDL-Eval~\cite{pyhdleval},
CreativEval~\cite{creativeval},
and ChipBench~\cite{chipbench}.

\noindent\textbf{Vision.}
We evaluate multimodal circuit and PCB understanding using
RFSchemBench~\cite{rfschembench},
PCB-Bench~\cite{pcbbench},
and CircuitSense~\cite{circuitsense}.

\noindent\textbf{Agentic.}
We evaluate agentic task execution using
Semiconductor Agent, PCB Agent,
and GameCraft-Bench~\cite{luo2026gamecraft}.

\subsection{Baselines}
We compare Micius-Flash with proprietary and open-weight models.
The proprietary baselines include
Gemini 3.0 Flash~\cite{gemini3flash},
Claude Sonnet 5.0~\cite{claudesonnet5},
and GPT-5.6 Sol~\cite{openai2026gpt56sol}.
The open-weight baselines include
DeepSeek-V4-Pro~\cite{deepseekv4pro},
Kimi K2.6~\cite{kimik26},
Muse Glimmer 30B~\cite{museglimmer30b},
Tencent Hy3~\cite{hy3},
MiniMax-M3~\cite{minimaxm3},
and Qwen3.8-27B~\cite{qwen3827b}.
All models are evaluated at maximum reasoning effort, except GPT-5.6, which uses the ``xhigh'' setting.

\input{tables/results}

\subsection{Results}

Table~\ref{tab:teacher-summary} presents the evaluation results of Micius-Flash and nine selected reference models across 25 benchmarks covering reasoning and knowledge, vision, coding, agentic capabilities, and general capabilities. Micius-Flash achieves leading results on selected circuit understanding and hardware-oriented coding benchmarks and ranks second on all three agentic engineering benchmarks among models with reported results. It also maintains competitive general-purpose reasoning and knowledge capabilities, demonstrating a broad capability profile across diverse technical domains.

\noindent\textbf{Reasoning and Knowledge.}
Micius-Flash scores 49.70\% on SemiBench, improving upon Qwen3.8-27B (42.51\%) by 7.19 points. On SuperGPQA, Micius-Flash consistently outperforms the base model across all four semiconductor-related subdomains. Specifically, it obtains 72.22\% on Circuits, 86.36\% on Electromagnetic Field, 84.00\% on Microelectronics, and 70.49\% on Physics, corresponding to improvements of 1.85, 2.27, 8.00, and 6.56 points, respectively. Notably, Micius-Flash achieves the highest score among the evaluated models on the Microelectronics subset. These results highlight Micius-Flash's capabilities in domain-specific knowledge understanding and technical reasoning across semiconductor and circuit engineering tasks.

\noindent\textbf{Vision.}
Micius-Flash demonstrates strong visual understanding capabilities in circuit and PCB engineering tasks. It achieves the highest scores among models with reported results on RFSchemBench (72.40\%) and CircuitSense (71.00\%), outperforming the strongest reference model on both benchmarks, Qwen3.8-27B (66.80\% and 69.80\%, respectively). On PCB-Bench, Micius-Flash ranks second among models with reported results with 80.60\%, behind GPT-5.6 Sol (86.96\%) and ahead of Qwen3.8-27B (80.00\%) and Kimi K2.6 (77.39\%).

\noindent\textbf{Coding.}
Micius-Flash achieves 86.80\% on VerilogEval-v2, 79.10\% on RTLLM-v2, 83.80\% on NetlistBench, and 65.80\% on PyHDL-Eval, demonstrating strong capabilities in hardware-oriented code generation and design. Notably, Micius-Flash ties GPT-5.6 Sol for the highest score on RTLLM-v2 and ranks second on NetlistBench, behind Kimi K2.6 (90.56\%) and ahead of Muse Glimmer (83.00\%) and Tencent Hy3 (82.84\%). Micius-Flash also ranks second on Rule2DRC with 48.30\%, behind GPT-5.6 Sol (75.50\%), and achieves 76.20\% on VHDLBench and 79.20\% on CreativEval. These results highlight Micius-Flash's capabilities in RTL generation, hardware design, and engineering-oriented code reasoning.

\noindent\textbf{Agentic Capabilities.}
Micius-Flash delivers strong performance on agentic engineering benchmarks, achieving 62.41\% on Semiconductor Agent, 83.30\% on PCB Agent, and 50.50\% on GameCraft-Bench. It ranks second on all three benchmarks among models with reported results. On Semiconductor Agent, Micius-Flash scores below GPT-5.6 Sol (87.25\%) while outperforming DeepSeek-V4-Pro (57.29\%) and Qwen3.8-27B (45.22\%). On PCB Agent, Micius-Flash ranks behind GPT-5.6 Sol (86.70\%) and ahead of DeepSeek-V4-Pro (80.00\%) and MiniMax-M3 (73.30\%). On GameCraft-Bench, Micius-Flash ranks behind GPT-5.6 Sol (60.50\%) and exceeds Qwen3.8-27B (32.10\%). These results highlight Micius-Flash's capabilities in multi-step engineering tasks and interactive development environments.

\noindent\textbf{General Capabilities.}
Beyond its specialized engineering capabilities, Micius-Flash maintains competitive general-purpose performance in mathematical reasoning and knowledge-intensive tasks. Micius-Flash achieves 87.90\% on HMMT, outperforming Qwen3.8-27B (81.82\%) and achieving comparable results to MiniMax-M3 and Tencent Hy3 (both 87.88\%). On MMLU-Pro, Micius-Flash scores 84.20\%, surpassing Qwen3.8-27B (83.60\%) and Muse Glimmer (83.50\%).

Taken together, these results demonstrate Micius-Flash's capabilities across semiconductor knowledge and reasoning, visual engineering understanding, hardware-oriented coding, and agentic engineering tasks. Micius-Flash achieves leading results on several specialized benchmarks and competitive results on general-purpose benchmarks, although its relative performance varies across tasks.

%% file: tables/results.tex
\begin{table}[!t]
\centering
\caption{Results across 25 benchmarks for nine selected reference models and Micius-Flash.}
\label{tab:teacher-summary}

\begingroup
\fontsize{7.2}{8.6}\selectfont
\setlength{\tabcolsep}{0.6pt}
\renewcommand{\arraystretch}{1.17}

\newcommand{\miciusSummaryHead}[1]{%
  \parbox[c][28pt][c]{\linewidth}{%
    \centering\fontsize{6.2}{7.3}\selectfont
    \bfseries\shortstack[c]{#1}}}

\newcommand{\miciusSummaryHeadCompact}[1]{%
  \parbox[c][28pt][c]{\linewidth}{%
    \centering\fontsize{6.2}{7.3}\selectfont
    \bfseries\scalebox{0.85}[1]{\shortstack[c]{#1}}}}

\newcommand{\miciusSummaryBand}[1]{%
  \rowcolor{seedblue!12}
  \multicolumn{11}{@{}l@{}}{%
    \rule[-2pt]{0pt}{13pt}\hspace{4pt}
    \textcolor{seedblue!65!black}{\textbf{#1}}}\\}

\begin{tabularx}{\linewidth}{%
@{}>{\raggedright\arraybackslash}p{97pt}|
*{9}{>{\centering\arraybackslash}X}|
>{\columncolor{seedblue!7}\centering\arraybackslash}X@{}}

\toprule

\multicolumn{1}{@{}c|}{\parbox[c][28pt][c]{97pt}{\centering\textbf{Benchmark}}}
& \miciusSummaryHead{Qwen\\3.8-27B}
& \miciusSummaryHead{Muse\\Glimmer}
& \miciusSummaryHead{MiniMax\\M3}
& \miciusSummaryHead{Tencent\\Hy3}
& \miciusSummaryHead{Kimi\\K2.6}
& \miciusSummaryHead{DeepSeek\\V4-Pro}
& \miciusSummaryHeadCompact{Gemini\\3.0 Flash}
& \miciusSummaryHeadCompact{Claude\\Sonnet 5.0}
& \miciusSummaryHead{GPT\\5.6 Sol}
& \textcolor{seedblue!65!black}{%
    \miciusSummaryHead{\textbf{Micius}\\\textbf{Flash}}}
\\

\midrule

\miciusSummaryBand{Reasoning and knowledge}

SemiBench
& 42.51 & 29.94 & 56.89 & 52.10 & 45.51 & 67.66 & 68.50 & 66.40 & 88.02 & 49.70 \\

SuperGPQA (Circuits)
& 70.37 & 63.89 & 61.11 & 73.15 & 67.59 & 73.15 & 72.55 & 67.86 & 75.93 & 72.22 \\

SuperGPQA (Emfield)
& 84.09 & 82.95 & 78.41 & 86.36 & 88.64 & 85.23 & 87.50 & 86.67 & 87.50 & 86.36 \\

SuperGPQA (Microelec)
& 76.00 & 70.00 & 82.00 & 78.00 & 82.00 & 82.00 & 78.12 & 79.31 & 78.00 & 84.00 \\

SuperGPQA (Physics)
& 63.93 & 63.93 & 63.93 & 70.49 & 72.13 & 70.49 & 85.37 & 72.22 & 65.57 & 70.49 \\

\miciusSummaryBand{Vision}

RFSchemBench
& 66.80 & 54.87 & 54.56 & - & 53.01 & - & - & - & 50.84 & 72.40 \\

PCB-Bench
& 80.00 & 72.17 & 77.39 & 47.83 & 77.39 & - & - & - & 86.96 & 80.60 \\

CircuitSense
& 69.80 & 52.07 & 41.42 & 26.04 & 49.70 & - & - & - & 60.36 & 71.00 \\

\miciusSummaryBand{Coding}

CVDP-static
& 43.70 & 39.40 & 57.95 & 59.60 & 50.99 & 57.28 & 57.00 & - & 65.80 & 58.90 \\

VerilogEval-v2
& 77.80 & 82.60 & 81.35 & 88.42 & 77.81 & 89.71 & 50.50 & 66.30 & 94.20 & 86.80 \\

RTLLM-v2
& 65.10 & 65.10 & 76.74 & 74.42 & 72.09 & 79.07 & 78.00 & 69.60 & 79.10 & 79.10 \\

ArchXBench
& 50.70 & 56.30 & 59.15 & 64.79 & 53.52 & 66.20 & 67.80 & 82.60 & 80.00 & 53.50 \\

NetlistBench
& 69.80 & 83.00 & 75.62 & 82.84 & 90.56 & 77.97 & 68.95 & 79.40 & 75.60 & 83.80 \\

VHDLBench
& 70.90 & 73.80 & 71.84 & 75.24 & 72.33 & 83.50 & 86.50 & 83.20 & 84.50 & 76.20 \\

Pluto
& 62.10 & 64.70 & 68.10 & 68.10 & 65.52 & 68.10 & 76.00 & 72.30 & 78.50 & 72.40 \\

NotSoTiny
& 14.70 & 12.70 & 14.83 & 14.49 & 12.57 & 17.36 & - & - & 18.80 & 16.10 \\

Rule2DRC
& 4.90 & 5.80 & 12.80 & 16.10 & 29.00 & 34.60 & 32.10 & 44.30 & 75.50 & 48.30 \\

PyHDL-Eval
& 19.60 & 49.80 & 35.22 & 41.47 & 46.53 & 53.67 & 67.10 & 60.60 & 74.50 & 65.80 \\

CreativEval
& 71.70 & 70.80 & 71.67 & 78.33 & 72.50 & 88.33 & 84.40 & 83.90 & 94.10 & 79.20 \\

ChipBench
& 39.10 & 39.90 & 46.62 & 50.38 & 51.13 & 50.38 & 47.40 & 57.00 & 46.60 & 50.40 \\

\miciusSummaryBand{Agentic capabilities}

Semiconductor Agent
& 45.22 & 28.80 & 39.20 & 33.80 & 22.70 & 57.29 & - & - & 87.25 & 62.41 \\

GameCraft-Bench
& 32.10 & - & - & - & 30.65 & 2.15 & - & - & 60.50 & 50.50 \\

PCB Agent
& 60.00 & - & 73.30 & 56.70 & 66.70 & 80.00 & - & - & 86.70 & 83.30 \\

\miciusSummaryBand{General Capabilities}

HMMT
& 81.82 & 57.58 & 87.88 & 87.88 & 75.76 & 90.91 & 78.80 & 90.00 & 93.94 & 87.90 \\

MMLU-Pro
& 83.60 & 83.50 & 83.27 & 87.20 & 87.30 & 88.40 & - & - & 88.90 & 84.20 \\

\bottomrule
\end{tabularx}
\endgroup
\vspace{3pt}
\begin{minipage}{\linewidth}
\fontsize{7.0}{8.4}\selectfont
\par\smallskip
\end{minipage}
\end{table}

%% file: sections/05_Micius_harness.tex
\section{Micius Harness}
\label{sec:Micius_harness}

\begin{figure}[!htb]
    \centering
    \includegraphics[width=\linewidth,height=\textheight,keepaspectratio]{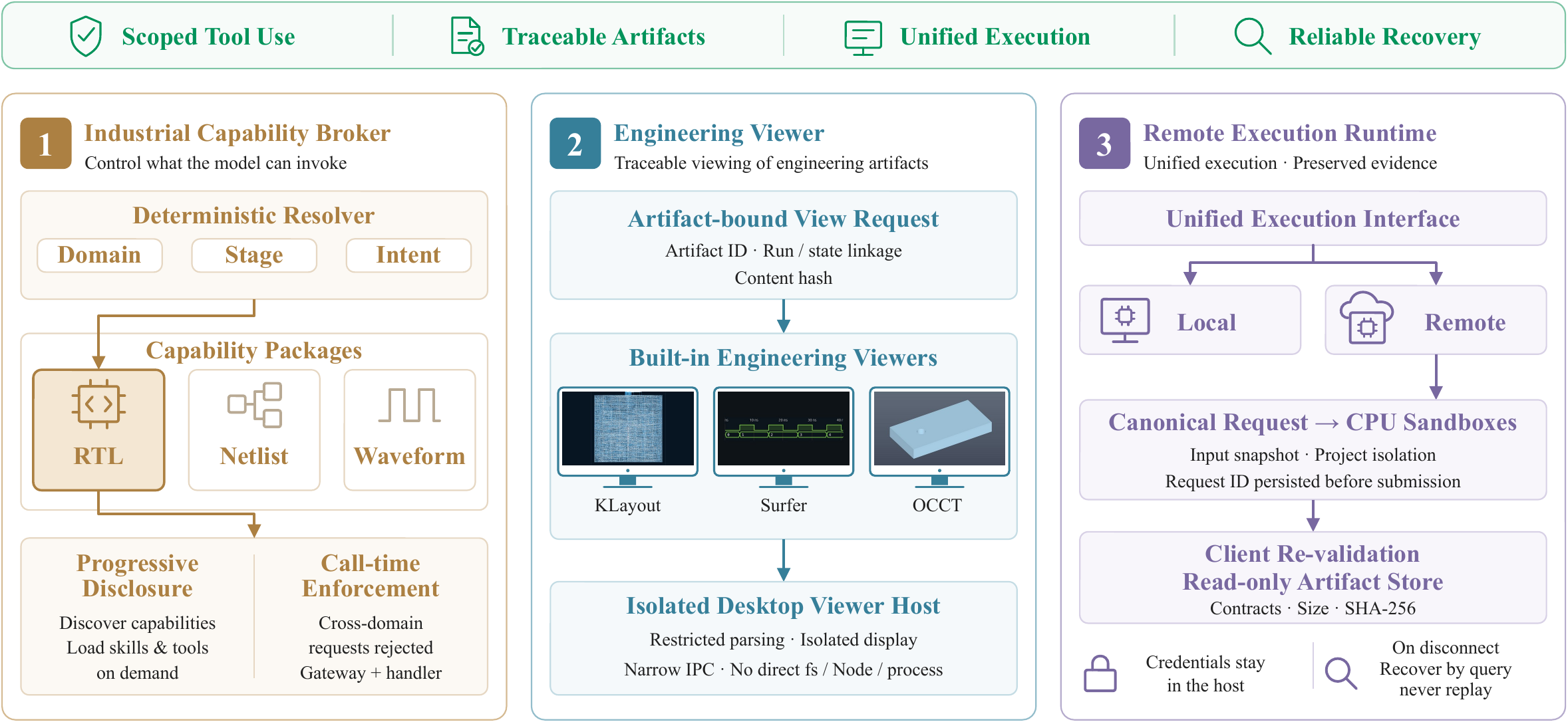}
    \caption{Micius Harness overview. The Capability Broker selects and enforces a scoped tool set; the Engineering Viewer provides artifact-bound inspection; and the Remote Execution Runtime provides local and remote execution, querying saved invocation identities to recover ambiguous calls while leaving unresolved outcomes unknown.}
    \label{fig:micius-harness}
\end{figure}

Micius Harness is an engineering workbench that connects Micius-Flash's domain knowledge, code generation, and tool-use capabilities to professional software and real project environments. Industrial workflows span stages with distinct tools and preconditions, and produce artifacts such as layouts, netlists, waveforms, and boards that require specialized inspection. Some workloads can exceed workstation resources, motivating remote execution while raising requirements for controlled data transfer and reliable failure recovery. As shown in Figure~\ref{fig:micius-harness}, three subsystems support these workflows: the Capability Broker scopes tool access, the Engineering Viewer supports artifact inspection, and the Remote Execution Runtime provides a shared interface for local and remote jobs.

The following describes the core functions and control mechanisms of these three subsystems:

\begin{itemize}
    \item \textbf{Deterministic Capability Scoping}
    Tool needs vary across engineering domains and workflow stages. The Capability Broker organizes tools as domain capabilities associated with stages such as RTL design, synthesis, physical design, and verification for chips, or schematic capture and board layout for PCBs. When domain and stage information is available, it filters registered capabilities by those fields, ranks candidates using task, tool, and artifact matches, and exposes the selected capabilities and tools to Micius. Canonical tool identifiers decouple capability definitions from provider-specific names. Selection is deterministic for a given registry and request; the trace records candidates, selections, and scope changes for later inspection. At invocation, the Runtime checks the tool allowlist and project and state bindings against the current request. Micius receives a task-relevant tool set, and the Runtime enforces that scope at each invocation.

    \item \textbf{Built-in Traceable Artifact Visualization.}
    Engineers need specialized views of design artifacts and a way to verify which file they are inspecting. The Engineering Viewer provides built-in viewers for supported chip layouts, synthesized netlists, simulation waveforms, and PCB boards and schematics. Project files selected for viewing are registered with artifact identities, and their content hashes are checked when opened. Views opened from recorded results retain links to their originating runs and check records where available. Viewer implementations apply format-specific input and resource checks, while the renderer accesses viewer resources through a restricted inter-process communication interface. Rendering an artifact or opening it in an external application does not run or change engineering checks. These mechanisms let engineers inspect supported artifacts while checking file content and, where recorded, its source.

    \item \textbf{Unified Remote Execution and Recovery.}
    Simulation, synthesis, and physical-design workloads can exceed an engineer's workstation resources. For projects configured for remote execution, Micius uses the same task interface as for local jobs, while the runtime handles remote transfer, result validation, and recovery. Selected files require explicit approval before upload, service credentials remain on the host side and outside the model process environment, and returned records and artifact sizes and hashes are validated before import. When a request's outcome is unclear, the client queries its saved invocation identity; unresolved outcomes remain unknown and are not resubmitted, avoiding duplicate execution.
\end{itemize}

%% file: sections/06_Case_study.tex
\section{Capabilities of Micius}
\label{sec:Case_studies}

To complement the preceding quantitative evaluations, we present six representative case studies illustrating how Micius-Flash applies its capabilities in practical scientific, engineering, and software development workflows. The cases span chip design, industrial software development, autonomous experimentation, PCB engineering, game development, and GPU kernel optimization. Across these diverse tasks, Micius-Flash translates high-level requirements into concrete implementations or experimental results, interacts with domain-specific tools, and iteratively refines its work based on execution feedback. For each case, we describe the task objectives, key execution steps, resulting outputs, and available validation evidence, highlighting how the model identifies errors, adapts its approach, and verifies its results through interaction with real or simulated tool environments. Together, these case studies provide a concrete view of how Micius-Flash combines domain knowledge, tool use, and iterative problem-solving to accomplish practical tasks across diverse technical domains.

\begin{figure}[!h]
    \centering
    \includegraphics[
        width=\textwidth,
        height=0.8\textheight,
        keepaspectratio
    ]{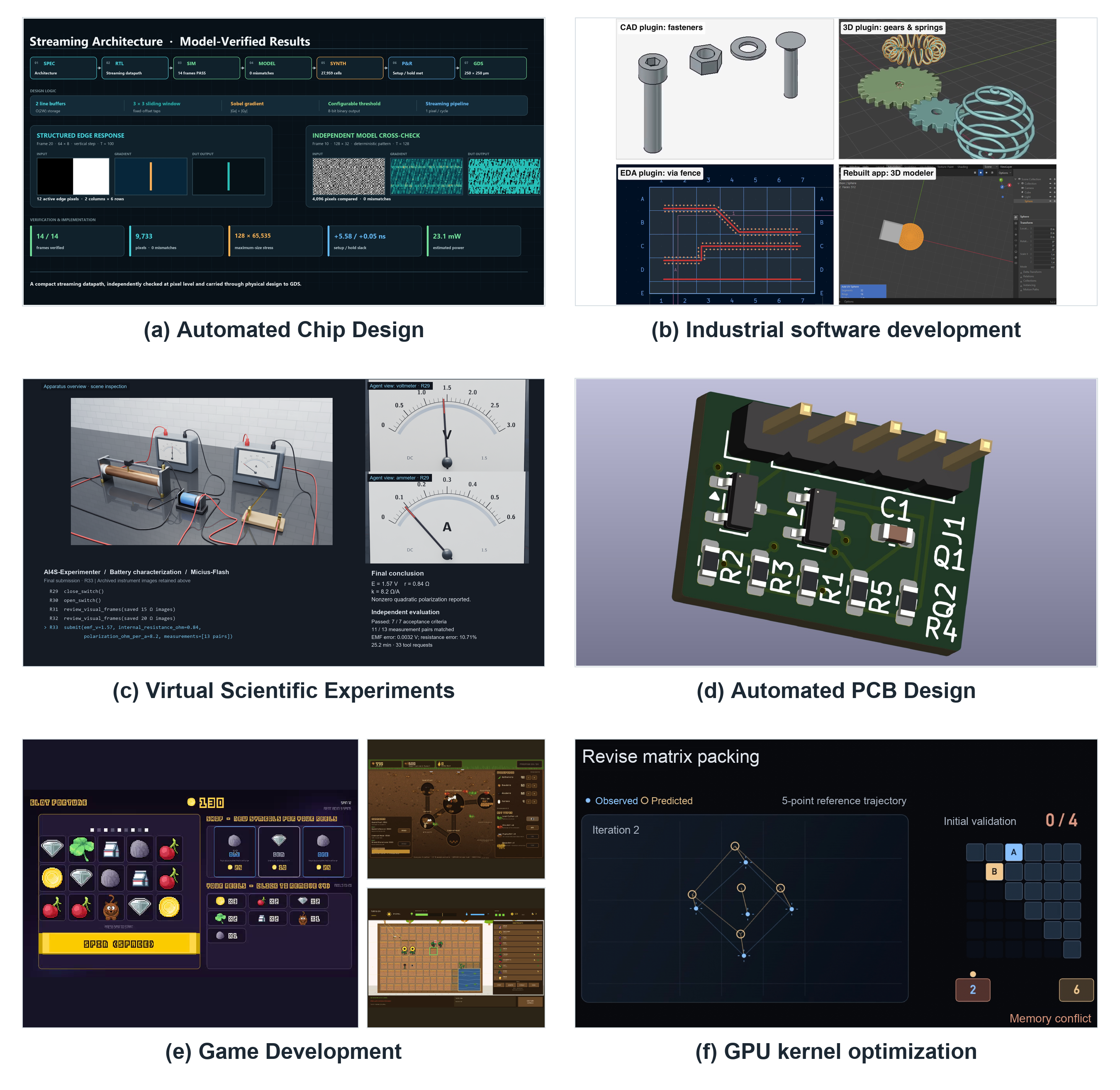}
    \caption{Representative scientific and engineering applications of Micius-Flash. (a) Automated design, verification, and physical implementation of a streaming Sobel edge detector. (b) Development of CAD, 3D modeling, and EDA plugins, together with reconstruction of a 3D modeling application. (c) Autonomous battery characterization through instrument operation and parameter estimation in a virtual laboratory. (d) 3D rendering of an automatically designed soft-latch PCB. (e) Playable Godot games with implemented interfaces and gameplay mechanics. (f) Visual feedback for diagnosing matrix-packing errors during GPU kernel optimization.}
    \label{fig:micius-capabilities}
\end{figure}

\noindent\textbf{Chip design agent.} Micius-Flash interacts directly with EDA tools to translate hardware design requirements into synthesizable RTL, verified implementations, and physical layouts. The model coordinates architecture design, functional verification, logic synthesis, and physical implementation, iteratively refining the design based on feedback from real engineering tools. In the Sobel real-time image filter case study, the model uses an iterative loop of research, design, experiment, analyze, and refine to take an 8-bit grayscale streaming edge detector from microarchitectural specification to physical signoff. The design pairs dual line buffers with a shift register array to maintain a sliding window over incoming pixels, computing Sobel gradients and applying configurable thresholding under cycle-accurate frame configuration, input handshaking, and output ordering control. Alongside synthesizable RTL, the model builds an independent golden reference model for differential verification and drives the backend flow, including logic synthesis, place and route, parasitic extraction, and post-route static timing analysis (STA), targeting the Nangate45/FreePDK45 process. When initial post-route STA exposed critical path violations caused by combinational logic depth and wire parasitics, the model inspected the timing reports, stepped back to the RTL to pipeline the arithmetic datapath, and realigned the control path latency. After tuning physical implementation parameters and passing regression tests, the design closed timing at 100 MHz; the model then added transistor-level LVS checks to deliver a fully verified GDS layout and signoff package, as shown in Figure~\ref{fig:micius-capabilities}a. Rather than acting as a single-pass code generator, the model uses EDA tool feedback to diagnose cross-layer bottlenecks and produce verifiable engineering artifacts. This approach extends naturally to architectural exploration, verification, and PPA optimization for processors and accelerators, as well as device model calibration, parameter exploration, and post-silicon test analysis.

\noindent\textbf{Industrial software agent.} Micius-Flash supports two industrial software development tasks: plugin development, which extends existing engineering applications with native functionality, and software replication, which rebuilds desktop applications from observed behavior without access to their source code. In both, it works directly against the real software to implement, test, and iteratively refine its output. In plugin development, given a one-paragraph requirement and the host's official documentation, Micius-Flash built a hole-pattern workbench for an open-source CAD system from scratch, supporting circular and linear hole patterns and hole reports. It located the headless command-line build inside the installed application, ran probe scripts there to verify hole placement, and, when the host rejected its package metadata, searched the host's binaries for the error message and read the bundled metadata files to recover the expected format. After six check runs inside the host, it completed the 376-line plugin in a single 33-minute session of 29 model calls; plugins built the same way for CAD, 3D modeling, and EDA applications are shown in Figure~\ref{fig:micius-capabilities}b. In software replication, Micius-Flash rebuilt the modeling workflow of a full-featured open-source 3D modeling suite as a standalone PySide6 application. It first operated the original like a user, opening menus, inspecting controls, and capturing reference views, and then, over ten build rounds, implemented the scene data model, mesh editing operations, an interactive 3D viewport, the main window layout, and object operations such as transforms, parenting, and modifiers. After each round, identical input sequences were replayed on both applications and the differences were fed back to Micius-Flash. The 18-hour run took about 1,800 model calls, and the resulting replica reproduces the original's default scene, interface layout, add menu, primitive creation, and material assignment under the same inputs. These cases show that Micius-Flash can both extend existing engineering software and rebuild functional applications through direct interaction with real software environments.

\noindent\textbf{Autonomous experimentation agent.} 
Micius-Flash conducts scientific experiments in simulated laboratory environments, directly operating instruments to collect measurements, analyze observations, and infer unknown physical parameters. In AI4S-Experimenter, a virtual laboratory benchmark that varies underlying parameters or introduces hidden anomalies while keeping task descriptions unchanged, the model must select measurement actions, interpret instrument readings, and draw conclusions from experimental evidence. The evaluation assesses both the consistency of reported measurements with instrument records and the accuracy of parameter estimates against simulator ground truth. In a battery characterization case, Micius-Flash operated a switch and rheostat, read analog voltmeter and ammeter displays, and collected measurements across five load settings to estimate the cell's electromotive force, internal resistance, and nonlinear polarization coefficient. Before submission, it revisited saved instrument images to verify its readings (Figure~\ref{fig:micius-capabilities}c). The experiment lasted 25.2 minutes and involved 33 tool requests. The estimated electromotive force had an absolute error of \(0.0032\,\mathrm{V}\), while the internal resistance had a relative error of \(10.71\%\), with all seven acceptance criteria satisfied. This case demonstrates Micius-Flash's ability to autonomously conduct multi-step experiments, interpret visual measurements, and derive quantitative conclusions grounded in experimental evidence.

\noindent\textbf{PCB engineering agent.} 
In a dedicated KiCad environment, Micius-Flash interacts directly with engineering tools to turn circuit requirements into native schematic and PCB projects and make local repairs to existing designs. The model constructs schematics, places footprints, and routes connections, using native electrical rule checks (ERC), design rule checks (DRC), netlists, and copper-connectivity verification to check design connections and specified nominal states. In the soft-latch case, Micius-Flash started from a blank schematic and a fixed board outline and produced a \(5\,\mathrm{V}\) switching circuit and a two-layer PCB with nine footprints (Figure~\ref{fig:micius-capabilities}d). The design uses an AO3401A P-channel MOSFET as a high-side switch, together with an MMBT3904 transistor circuit, and exposes a five-pin VIN/GND/VOUT/SW/HOLD interface for startup and power retention: grounding SW enables the output, an asserted HOLD signal maintains power after the startup switch is released, and releasing SW while deasserting HOLD turns the output off. Other designs include a 13-component LED latch based on a 74HC74 flip-flop, supporting button-triggered set, hold, and reset operations, and a seven-component single-cell battery charger using an LTC4054ES5-4.2 with a \(10\,\mathrm{k}\Omega\) programming resistor for a nominal charging current of \(100\,\mathrm{mA}\). In a thumb-keyboard repair task, Micius-Flash reconnected connector pin J1.1 to \texttt{COL\_5} while preserving the existing 20-component circuit. The repaired netlist confirmed the connector mapping and six distinct key row--column pairs, and the designs were delivered as native KiCad project files that can be reopened and inspected.

\noindent\textbf{Game development agent.} 
Micius-Flash interacts directly with the Godot development environment to turn natural-language gameplay and interface requirements into editable, playable game projects. The model first identifies gameplay mechanics, interaction rules, and state transitions and implements the corresponding scenes, scripts, and UI components. It then runs the game, inspects visual feedback, and replays interactions to locate and correct functional or interface errors. Three cases from GameCraft-Bench illustrate this process. In \textit{Slot Fortune}, Micius-Flash integrated the shop, currency, and reel inventory systems so that purchases update both the player's balance and inventory. In \textit{Idle Ant Empire}, the model inspected control coordinates to locate a layout error in the worker-assignment interface, revised the layout logic, and replayed the interaction to check the repair. In \textit{Garden Ecosystem Keeper}, it implemented turn-based environmental changes and crop growth, producing observable state progression across successive turns. The resulting native Godot projects retain editable scenes, scripts, and UI components and can be reopened, run, and further developed; accompanying execution records and screenshots show the corresponding interactions and state transitions (Figure~\ref{fig:micius-capabilities}e).

\noindent\textbf{GPU kernel optimization.} 
Micius-Flash uses tools and execution feedback to implement, debug, and optimize GPU kernels, reducing computational and scheduling overhead while preserving numerical correctness. In the Perspective-n-Point (PnP) pose refinement case (Figure~\ref{fig:micius-capabilities}f), the model implements Levenberg–Marquardt (LM) iterations to jointly optimize rotation and translation, aligning projected 3D points with their 2D observations. Guided by validation feedback, Micius-Flash identifies and corrects misplaced elements in the packed derivative matrix, then fuses projection, derivative computation, reduction, and linear system solving into a single GPU kernel that executes the full iterative procedure. For a public test case comprising four groups of 64 points each, this reduces the number of kernel launches from 1,322 in the GPU eager reference implementation to one. The final implementation passes all 12 independent validation tests and achieves a geometric mean speedup of 174.8× over that reference. This case demonstrates how Micius-Flash combines numerical algorithm understanding, error diagnosis, and kernel fusion to deliver a validated GPU implementation for iterative workloads dominated by fine-grained operator calls.

%% file: sections/07_conclusion.tex
\FloatBarrier
\section{Conclusion}

We present Micius-Flash, a 27B model with native multimodal capabilities for engineering, focusing primarily on semiconductor applications. By combining a knowledge-tree-driven data flywheel, expert training, and multi-teacher on-policy distillation, we bring complementary engineering capabilities into a unified model. Evaluations of Micius-Flash demonstrate gains over a same-size baseline in hardware coding and semiconductor agentic tasks. Through Micius Harness, the model interacts with specialized tools and execution environments, using feedback to plan, debug, and iteratively verify solutions across engineering workflows. 
In the future, we plan to explore more comprehensive industrial scenarios and scale Micius to larger model capacities, moving toward engineering agents that can reliably handle more complex tasks.

\label{sec:conclusion}

%% file: sections/08_Contributions.tex

\section*{Contributions}
\label{contributions}
\noindent\textbf{Core Contributors}


\noindent
Hongcheng Gao$^{\dagger}$$^{*}$, Weiheng Li, Yipeng Wei, Yu Lei, Shuo Sun, Yi Li, Yajiao Wang

\vspace{0.4em}

\noindent\textbf{Contributors}


\noindent
Jie Luo, Jianwei Wen, Xinyi Zhou, Hongda Li, Yibing Zhang, Henghui Sun, Yufan Zhou, Tianfang Hao, Naihao Xue, Hailong Qu, Jingjing Zhou, Peng Xu, Yongzheng Wei, Yichen Zhu, Ziying Wang, Yibo Wang, Jianshu Zeng, Yuhang Lei, Zeqiang Wang

\vspace{2em}

\noindent\rule{0.38\textwidth}{0.4pt}

{\footnotesize
\noindent $^{*}$Team Lead.

\noindent $^{\dagger}$Corresponding Author:
Hongcheng Gao \textless gaohongcheng2000@gmail.com\textgreater.
}

\clearpage